\documentclass[aps,twocolumn,nopacs,preprintnumbers]{revtex4}

\usepackage{graphicx, fancyhdr, amsfonts, amsmath, amssymb, hyperref, glossaries, siunitx}
\hypersetup{
    colorlinks=true,
    linkcolor=blue,
    filecolor=magenta,      
    urlcolor=cyan,
    pdftitle={Overleaf Example},
    pdfpagemode=FullScreen,
    }
\usepackage[T1]{fontenc}
\usepackage{xr}
\usepackage{color,soul}
\usepackage{physics}
\newcommand{\subtxt}[2]{\ensuremath{#1_\mathrm{#2}}}

\newacronym{dqd}{DQD}{double quantum dot}
\newacronym{qd}{QD}{quantum dot}
\newacronym{qw}{QW}{quantum well}
\newacronym{soi}{SOI}{spin--orbit interaction}
\newacronym{hh}{HH}{heavy-hole}
\newacronym{lh}{LH}{light-hole}
\newacronym[\glslongpluralkey={two dimensional hole gases}]{2dhg}{2DHG}{two dimensional hole gas}
\newacronym{haadf}{HAADF}{high-angle annular dark-field}
\newacronym{pcb}{PCB}{Printed Circuit Board}
\newacronym{vna}{VNA}{Vector Network Analyzer}
\newacronym{hemt}{HEMT}{High Electron Mobility Transistor}

\newacronym{hs}{HS}{heterostructure}
\newacronym{sshh}{SSHH}{superconductor/semiconductor hybrid heterostructure}
\newacronym{sige}{SiGe}{silicon-germanium}
\newacronym{ge}{Ge}{germanium}
\newacronym{si}{Si}{silicon}
\newacronym{pt}{Pt}{platinum}
\newacronym{al}{Al}{aluminum}
\newacronym{nb}{Nb}{Niobium}
\newacronym{cu}{Cu}{copper}

\newacronym{hf}{HF}{hydrofluoric acid}
\newacronym{rt}{RT}{Room Temperature}
\newacronym{bcs}{BCS}{Bardeen–Cooper–Schrieffer}
\newacronym{sc}{SC}{Superconductor}

\newacronym{sem}{SEM}{scanning electron microscopy}
\newacronym{tem}{TEM}{transmission electron microscopy}
\newacronym{cvd}{CVD}{chemical vapor deposition}
\newacronym{ald}{ALD}{atomic layer depostion}
\newacronym{rie}{RIE}{reactive ion etching}
\newacronym{sdh}{SdH}{Shubnikov–de Haas oscillations}
\newacronym{dhg}{2DHG}{two-dimensional hole gas}
\newacronym{edx}{EDX}{energy dispersive X-Ray}
\newacronym{icp}{ICP}{inductively coupled plasma}
\newacronym{tlf}{TLF}{two-level fluctuators}
\newacronym{flp}{FLP}{Fermi-level pinning}
\newacronym{vb}{VB}{valence band}
\newacronym{fl}{E$_F$}{Fermi-level}
\newacronym{alo}{Al$_2$O$_3$}{aluminum oxide}
\newacronym{sin}{SiN$_x$}{silicon nitride}
\newacronym{eb}{E$_b$}{thermal-activation energy}
\newacronym{np}{n$_p$}{percolation density}
\newacronym{cnl}{CNL}{charge neutrality level}
\newacronym{mos}{MOS}{metal-oxide semiconductor}
\newacronym{str}{s}{strained}
\newacronym{cptr}{CPTR}{coplanar transmission line resonator}
\newacronym{tl}{TL}{transmission line}
\newacronym{tls}{TLS}{two-level systems}
\newacronym{qp}{QP}{quasiparticle}
\newacronym{qps}{QPs}{quasiparticles}
\newacronym{dr}{DR}{dilution refrigerator}
\newacronym{mbe}{MBE}{molecular-beam epitaxy}
\newacronym{squid}{SQUID}{Superconducting Quantum Interference Device}
\newacronym{jj}{JJ}{Josephson junction}
\newacronym{cp}{CP}{Cooper pair}
\newacronym{cd}{CD}{Coulomb diamond}
\newacronym{fwhm}{FWHM}{Full Width at Half Maximum}
\newacronym{dc}{DC}{Direct-current}
\newacronym{rf}{RF}{Radio-frequency}
\newacronym{abs}{ABS}{Andreev-Bound-State}
\newacronym{rcsj}{RCSJ}{Resistively and Capacitively Shunted Junction}

\newacronym{qi}{\subtxt{Q}{i}}{internal quality factor}
\newacronym{qed}{QED}{Quantum Electrodynamcs}

\DeclareSIUnit\torr{Torr}
\DeclareSIUnit\evolt{eV}
\DeclareSIUnit\sccm{sccm}
\DeclareSIUnit{\square}{\ensuremath{\Box}} 

\DeclareSIUnit\dbm{dBm}
\DeclareSIUnit\degreekelvin{K}

\begin{document}

\title{Multi-level $\pi$-junction in a proximitized Ge/SiGe quantum dot probed
by an on-chip superconducting microwave resonator}
\author{Luigi Ruggiero\textsuperscript{1}}
\author{Vera Jo Weibel\textsuperscript{1}}
\author{Pauline Drexler\textsuperscript{2}}
\author{Carlo Ciaccia\textsuperscript{1}}
\author{Christian Olsen\textsuperscript{1}}
\author{Dominique Bougeard\textsuperscript{2}}
\author{Christian Schönenberger\textsuperscript{3, 4}}
\author{Andrea Hofmann\textsuperscript{1, 3, *}}

\begin{abstract} 
\begin{center}
\textsuperscript{1}  University of Basel, Klingelbergstrasse 82, 4056 Basel, Switzerland \\
\textsuperscript{2} University of Regensburg, Universitätsstraße 31, 93053 Regensburg, Germany \\
\textsuperscript{3} Swiss Nanoscience Institute, Klingelbergstrasse 82, 4056 Basel, Switzerland\\
\textsuperscript{4} YQuantum, Parkstrasse 1, 5234 Villigen, Switzerland\\
\textsuperscript{*} Corresponding author. Email: andrea.hofmann@unibas.ch\\
\vspace{1em}

\date{\today}
\end{center}

Using on-chip microwave measurements, we investigate multilevel $\pi$-junctions formed by proximitized \gls{qd} in a \gls{ge}/\gls{sige} heterostructure. In the multilevel regime, where several \gls{qd} orbitals contribute simultaneously to superconducting transport, the Josephson ground state is no longer determined solely by the occupation of a single orbital. By combining DC transport and microwave techniques, we identify the qualitative signatures of multilevel $\pi$-junctions in both their gate-voltage dependence and microwave response. In particular, we observe combinations phase transitions that are sharp or smooth in gate voltage and which exhibit distinct inductive and dissipative signatures. Such multilevel Josephson transport has previously been observed primarily in exceptionally clean systems such as carbon nanotubes. Our results establish proximitized \gls{ge} as a platform for investigating hybrid superconductor/semiconductor physics and demonstrate the integration of gate-defined superconducting quantum devices with high-quality on-chip microwave resonators.

\end{abstract}

\maketitle 

\section{Introduction}
Hybrid semiconductor–superconductor devices provide a versatile platform for exploring the interplay of superconductivity, Coulomb interaction, and quantum confinement.\glspl{qd} coupled to superconducting reservoirs are of particular interest because they yield \glspl{jj} whose ground state can be controlled electrically \cite{defranceschi_hybrid_2010}. These junctions have been studied extensively in the single-level limit \cite{cleuziou_carbon_2006, eichler_evenodd_2007, pillet_andreev_2010, lee_spinresolved_2014a}, where transport is governed by a single orbital, as well as in the limit of metallic islands involving a large number of orbitals \cite{aumentado_nonequilibrium_2004, echternach_progress_2009}. Between these two extremes lies the multi-level regime, where the single-particle energy remains relevant but several orbitals simultaneously contribute to transport \cite{shimizu_multilevel_1998, vanheck_conductance_2016}. Experimental studies of this regime are scarce \cite{vandam_supercurrent_2006, delagrange_0_2018} and have so far relied on transport measurements, whereas microwave techniques provide additional access \cite{bargerbos_spectroscopy_2023} to the inductive and dissipative response of the junction.

Planar \gls{ge}/\gls{sige} \glspl{qw} have recently emerged as a platform that combines gate-defined \glspl{qd} \cite{hofmann_assessing_2019a, nigro_demonstration_2024b, hendrickx_fourqubit_2021b, jirovec_singlettriplet_2021b}, the superconducting proximity effect \cite{hendrickx_gatecontrolled_2018, valentini_parityconserving_2024a, vigneau_germanium_2019b, aggarwal_enhancement_2021b}, and superconducting microwave resonators \cite{ruggiero_high-quality_2025, nigro_demonstration_2024b}. Recent advances in \gls{ge}-based hybrid superconductor-semiconductor devices have enabled proximitized \glspl{qd} \cite{lakic_quantum_2025, fabris_granular_2026}, gate-tunable transmon qubits \cite{sagi_gate_2024b, kiyooka_gatemon_2025a} and high-transparency \glspl{abs} \cite{hinderling_direct_2024a, kate_finite_2025}. These developments now make it possible to study the phase-dependent transport through proximitized quantum dots within a planar semiconductor platform using combined \gls{dc} and \gls{rf} methods. While control over the superconducting phase \subtxt{\varphi}{0}, which determines the phase minimizing the junction ground-state energy, has previously been demonstrated using \glspl{qd} in nanowires \cite{vandam_supercurrent_2006,sahu_ground-state_2024} and carbon nanotubes \cite{delagrange_0_2018, debbarma_josephson_2023}, implementations in planar material systems have largely relied on superconducting islands \cite{aumentado_nonequilibrium_2004, mannila_detecting_2019,hinderling_flipchipbased_2024}.

In a \gls{sc}-\gls{qd}-\gls{sc}, the interplay between the pairing induced by the \gls{sc} and the electron-electron interaction provided by the confinement into a \gls{qd} defines the parity of the \gls{qd} ground state. The superconducting phase offset \subtxt{\varphi}{0}, however, depends not only on the ground state parity but also on the number and nature of excited states participating in the transport across the junction. To determine \subtxt{\varphi}{0}, we embed the junction in a \gls{squid} to flux tune the superconducting phase which the \glspl{cp} pick up when being transported across the \gls{sc}-\gls{qd}-\gls{sc} junction. We can induce sharp transitions from \subtxt{\varphi}{0} = $0$ (positive supercurrent) to $\pi$ (negative supercurrent) by changing the \gls{qd} occupancy and thus the current-phase relation of the ground state. We also demonstrate a gradual inversion of the supercurrent occurring at fixed charge on the \gls{qd}, when multiple \gls{qd} levels participate in the transport \cite{shimizu_multilevel_1998, vandam_supercurrent_2006, delagrange_0_2018}.
\begin{figure*}[ht!]
        \centering
        \includegraphics[width = 1.0\textwidth]{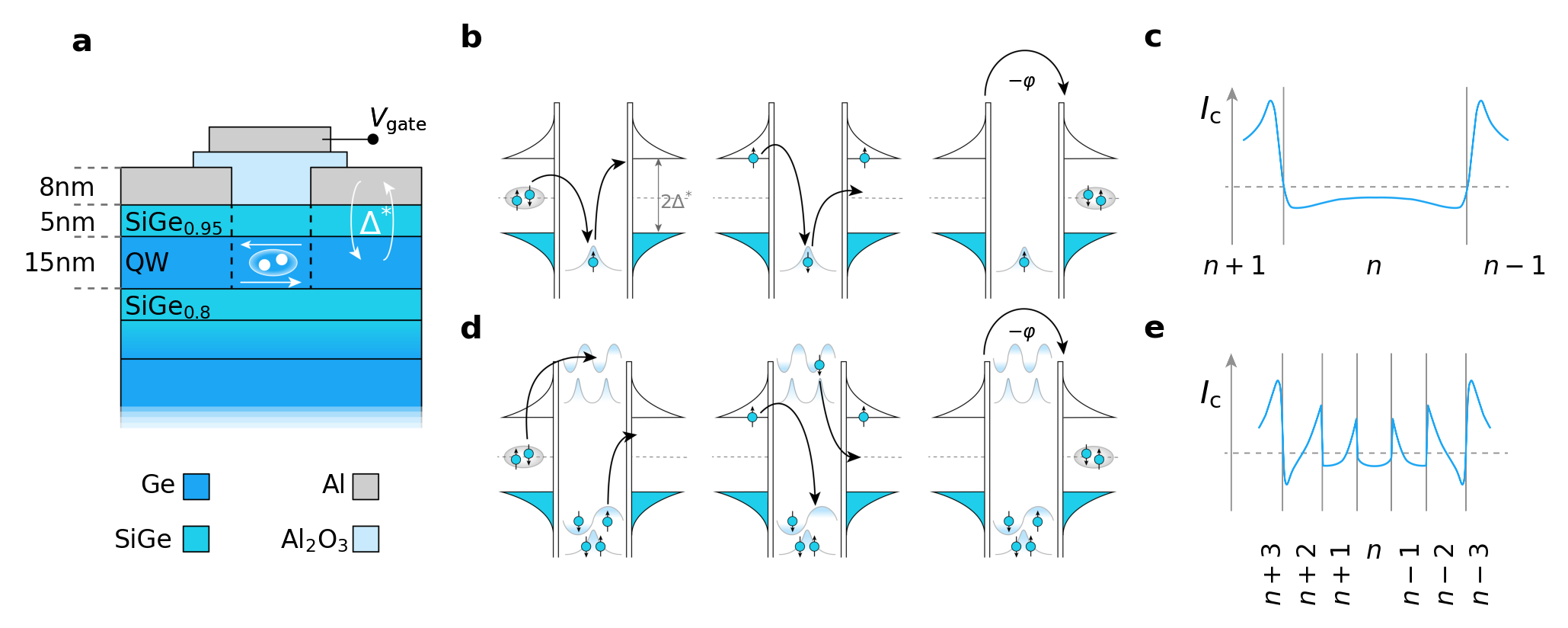}
        \caption{\textbf{Sketch of the heterostructure and the $\pi$-junction mechanisms in a proximitized \gls{qd}}. (a) Sketch of the hybrid \gls{al} - \gls{ge}/\gls{sige} heterostructure. The high \gls{ge} concentration in the top barrier and the \SI{5}{\nano\meter} thickness enhance the proximity of the \gls{al} into the \gls{qw}. (b) \glspl{cp} transport across a \gls{sc}-Doublet-\gls{sc}. Coulomb repulsion does not favor the simultaneous transport of both charges. The process stays coherent through the permutation of the spins of the \gls{cp}, resulting in a negative contribution to the total current. (c) Schematic of the critical current sign as a function of dot occupation if no higher order co-tunneling is involved. Dashed line marks a zero line. (d) \glspl{cp} transport across a \gls{sc}-\gls{qd}-\gls{sc} with higher order co-tunneling processes involved. Permutation of the spins is mediated by higher order \gls{qd} states, resulting in a negative contribution to the total current. (e) Schematic of the critical current sign as a function of dot occupation with higher order co-tunneling involved. Dashed line marks a zero line.}
        \label{fig:HS}
\end{figure*}
Our experiments reveal an induced superconducting gap of $\Delta^{*} \simeq$ \SI{240}{\micro\electronvolt}, higher than what has been reported so far for superconducting devices on \gls{ge} heterostructures \cite{valentini_parityconserving_2024a, lakic_quantum_2025, hinderling_direct_2024a}.
This, together with the field resilient \gls{cptr} \cite{ruggiero_high-quality_2025} paves the way towards exploring superconductivity, large \gls{soi} and sizeable Zeeman energies in a single device and at fast time-scales.

Our devices are fabricated out of a shallow \SI{15}{\nano\meter} \gls{ge}/\gls{sige} \gls{qw} proximitized by an in-situ grown nominally \SI{8}{\nano\meter} thick \gls{al} thin film \cite{ruggiero_high-quality_2025}. A detailed sketch of the stack is shown in \autoref{fig:HS}(a). The \gls{hs} is \gls{mbe} grown directly on top of a commercial \gls{ge} substrate. It has been demonstrated that this approach reduces the dislocation density at the interface between the substrate and the \gls{hs} \cite{nigro_demonstration_2024b} and allows for the implementation of high quality superconducting \gls{cptr} \cite{ruggiero_high-quality_2025}. The \gls{2dhg} is confined in the \gls{ge} layer, in between two strongly asymmetric \gls{sige} barriers of \SI{80}{\percent} and \SI{95}{\percent} \gls{ge} content for the bottom and top barrier, respectively. The asymmetric \gls{ge} concentration together with the \SI{5}{\nano\meter} thin top barrier aim to enhance the proximity effect and thus the induced superconducting gap $\Delta^{*}$.\\
The \gls{jj}, a weak link between two superconducting current leads, is a basic building block of a superconducting hybrid device. To lowest order, the supercurrent across a \gls{jj} follows a sinusoidal current-phase relation, $I(\phi) \propto \sin{(\varphi + \subtxt{\varphi}{0})}$, where $\varphi$ is phase difference across the superconducting leads. In general the \gls{jj} energy is minimized at a phase \subtxt{\varphi}{0}$=0, \pi$, if time reversal symmetry holds. In a \gls{sc} - \gls{qd} - \gls{sc} type of \gls{jj}, where a superconducting current is carried by discrete \gls{qd} levels, the phase offset is determined by the relative size of the competing interactions. While the Coulomb interaction, characterized by the charging energy $U$, enforces sequential tunneling through the \gls{qd}, the superconducting correlations influenced by the induced gap $\Delta^*$ and the tunnel coupling $\Gamma$ to the leads, drives coherent transport of \glspl{cp} across the junction. Low \gls{qd} excitation energies $\delta$ facilitate co-tunneling events. \\
\begin{figure*}[ht!]
        \centering
        \includegraphics[width = 1.0\textwidth]{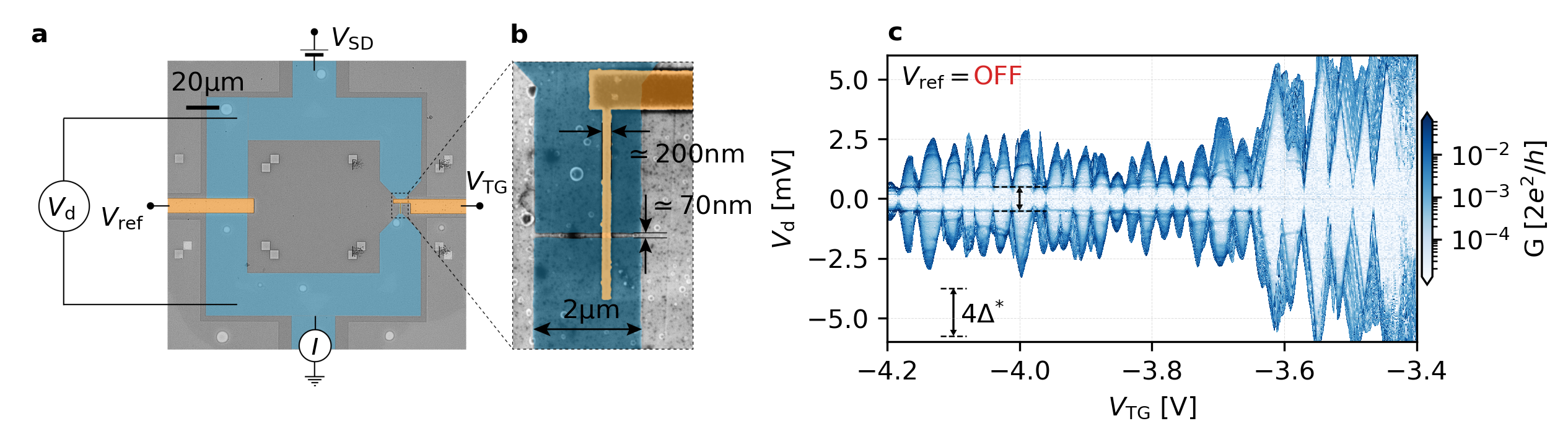}
        \caption{\textbf{DC-SQUID setup and single channel transport}.(a) Optical image of the \gls{dc}-\gls{squid} device with sketched relevant sizes and setup. (b) Zoomed-in \gls{sem} image of the \gls{jj} region of a test device. (c) Single channel conductance as a function of \subtxt{V}{TG}. The diamond pattern indicates the formation of a \gls{qd} persisting for a wide range of gate voltages. Outside the CD the missing data are caused by the residual resistance of the \gls{dc} line. The enhanced tunneling marked by black arrows match with an induced superconducting gap $\Delta^{*} \simeq$ \SI{240}{\micro\electronvolt}.}
        \label{fig:fig2}
\end{figure*}
When $\Delta^* \sim \Gamma  \sim U << \delta$, the \gls{qd} ground state is either a singlet or a doublet \cite{lakic_quantum_2025}. Tunneling through a single \gls{qd} level is favored, and the phase offset of the \gls{sc} - \gls{qd} - \gls{sc} junction is determined by the spin multiplicity of the \gls{qd} ground state. This is the single-level regime. On the other hand, when $\delta < U$, also co-tunneling processes to different orbitals strongly affect the transport, which puts the junction in a multi-level regime.

\autoref{fig:HS}(b) shows the case of a single-level $\pi$-junction, where the $\pi$ shift is mediated by spin-exchange processes during tunneling through a singly occupied \gls{qd} level forming a spin-degenerate doublet ground state. The \gls{cp} is transported from left to right across the \gls{qd} by a $\ket{\uparrow}$ tunneling to the right, while $\ket{\downarrow}$ occupies the \gls{qd} in a meta-stable state. $\ket{\downarrow}$ recombines with $\ket{\uparrow}$ in the right lead  to form a \gls{cp} while the state inside the \gls{qd} is again filled with a $\ket{\uparrow}$ from the left lead. On the other hand, if the \gls{qd} is empty or doubly occupied, the ground state is a spin singlet, and the tunneling of the \gls{cp} occurs without phase accumulation. In this regime, the \gls{qd} occupation and the according ground state multiplicity directly determine the phase offset to take the values $0$ or $\pi$, which result in different contribution to the total current as sketched in \autoref{fig:HS}(c). In contrast, a multi-level $\pi$-junction describes a regime where $\delta$ is small, and permutation of the spins also occur due to high order co-tunneling processes, this time involving several \gls{qd} orbitals, as illustrated in \autoref{fig:HS}(d). The multi-level process is not directly dependent on the parity of the charges in the \gls{qd} and the probability of a $\pi$ shift is non-zero for both odd and even occupation \cite{shimizu_multilevel_1998}. In this regime, the 0-$\pi$ transition is either sharp or smooth. While a sharp transition denotes a change of occupation in the \gls{qd}, the smooth transition purely depends on the involved orbitals. This is illustrated in \autoref{fig:HS}(e), where we sketch the \subtxt{I}{c} behavior as a function of \gls{qd} occupation for a situation with four spin-degenerate \gls{qd} states with a small energy spacing $\delta$ compared to $\Delta$ and the thermal energy 
\cite{shimizu_multilevel_1998}. At charge $n$, two of these states are occupied and two are empty, determining an electron-hole symmetry point for this set of (nearly-)degenerate states. At higher or lower charge occupation, smooth transitions occur due to the weight with which the transport is electron-like or hole-like with respect to the given charge on the \gls{qd}. In the simple case of one spin-degenerate state, the picture again simplifies to the single-level regime shown in \autoref{fig:HS}(c). 

\section{Results}
\label{sec:results}

\subsection{DC - SQUID}
We present \gls{dc} measurements of a $\pi$ - junction formed in a \gls{sc} - \gls{qd} - \gls{sc} type of \gls{jj}. As shown in the false colored optical image in \autoref{fig:fig2}(a), the \gls{jj} consists of a gate defined $\sim$ \SI{200}{\nano\meter} wide, $\sim$ \SI{70}{\nano\meter} long channel, tuned by \subtxt{V}{TG} and is embedded in an asymmetric \gls{squid} with an etch defined \SI{25}{\micro\meter} wide, $\sim$ \SI{90}{\nano\meter} long reference \gls{jj}, tuned by the voltage \subtxt{V}{ref}. We refer to the critical and switching current of the channel and reference junction by \subtxt{I}{c/sw,ref} and \subtxt{I}{c/sw}, respectively. We use \subtxt{I}{sw} to refer to the observed current inducing transitions from the superconducting to the normal state, while we use \subtxt{I}{c} in our formulae. We define the normalized external magnetic flux \subtxt{\phi}{ext}$=\varphi=2\pi\subtxt{\Phi}{ext}/\subtxt{\Phi}{0}$, where \subtxt{\Phi}{ext} is the external flux, $\subtxt{\Phi}{0}\approx$ \SI{2.067}{\weber} the flux quantum.
\\
In \autoref{fig:fig2}(c), the reference \gls{jj} was turned off by setting $\subtxt{V}{ref}=$\SI{0}{\volt} to focus on one channel. We show the differential conductance as a function of gate voltage \subtxt{V}{TG} and measured voltage drop \subtxt{V}{d} across the channel. We observe clear \gls{cd} patterns across a wide range of gate voltage, indicating the stable formation and tunability of a \gls{qd}. The formation of a \gls{qd} by means of a single gate is a consequence of the operation point at low density together with the small effective size of the gated channel. The shrinking \glspl{cd} and the emergence of co-tunnelling features indicate that not only the occupation of the \gls{qd} but also the tunnel coupling to the leads is tuned by \subtxt{V}{TG}. The Coulomb interaction dominates in this regime, the \gls{sc} proximity effect is clearly visible by the gap between the cusps of the \glspl{cd} and the enhanced conduction at a bias energy of $2\Delta^{*}$, as indicated with a black arrow. We estimate an induced gap of $\Delta^{*} \simeq$\ \SI{240}{\micro\electronvolt}, higher than what has been reported so far for similar material combinations \cite{valentini_parityconserving_2024a}.\\
\begin{figure*}[ht!]
        \centering
        \includegraphics[width = 1.0\textwidth]{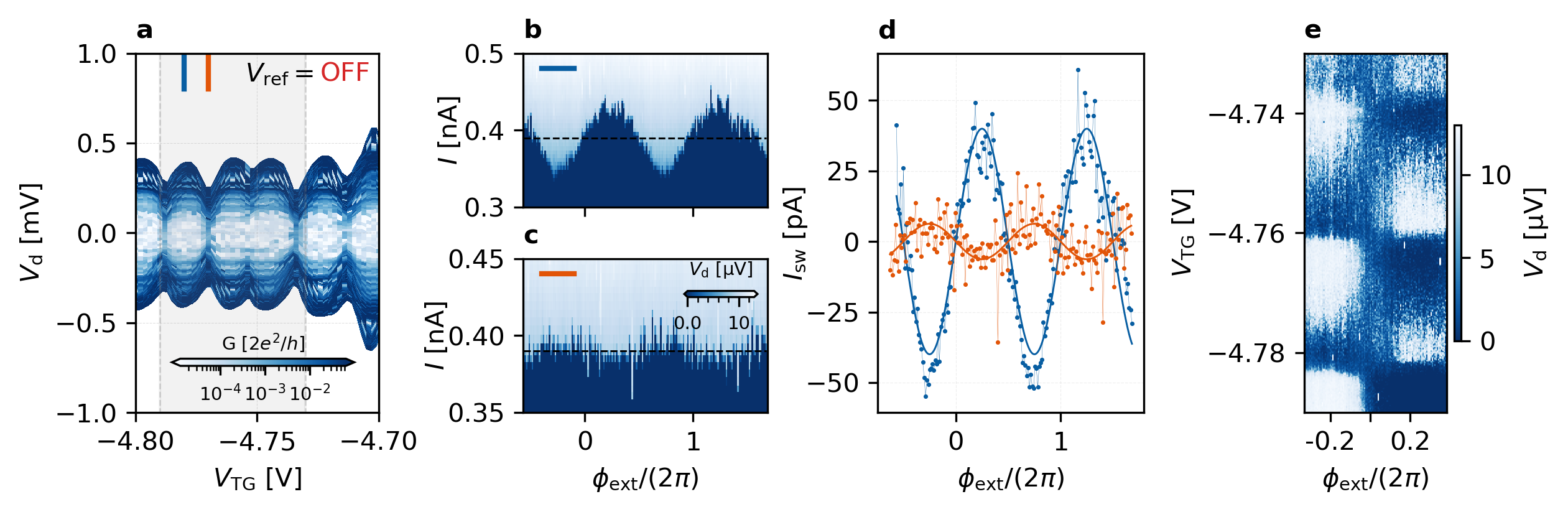}
        \caption{\textbf{0-$\pi$ transition in a DC-SQUID}. (a) Single channel conductance as a function of \subtxt{V}{TG}. Higher tunnel coupling to the leads results in a reduced the charging energy. (b-c) \gls{squid}-oscillation for two different \subtxt{V}{TG} voltages marked in (a). (d) Extracted \subtxt{I}{sw} (dots) from the maps in (b), with superimposed a sinusoidal fit (solid line). The two traces are $\pi$ shifted with respect to each other. (e) Fixed bias map \subtxt{I}{} $=$ \SI{0.39}{\nano\ampere} marked as a dashed line in (b), \subtxt{V}{TG} range highlighted as gray shaded region in (a). While tuning the occupation of the \gls{qd} the device undergoes multiple 0 to $\pi$ transitions.} 
        \label{fig:fig3}
\end{figure*}
In order to move into a transport regime where the Coulomb interaction and superconducting correlations become more comparable, we open the tunnel barriers by setting \subtxt{V}{TG} to more negative voltages as shown in \autoref{fig:fig3}. Here, the \glspl{cd} are less sharp and the increased tunnel coupling gives rise to a measurable supercurrent. At this point we switch on the reference junction by setting \subtxt{V}{ref} such that \subtxt{I}{sw ,ref} $\simeq \SI{0.39}{\nano\ampere}$, allowing us to control the phase drop over the \gls{jj} via the external flux \subtxt{\phi}{ext} threading the \gls{squid}. We sweep \subtxt{\phi}{ext} as shown in \autoref{fig:fig3}(b-c) for the two values of \subtxt{V}{TG} marked in \autoref{fig:fig3}(a). \autoref{fig:fig3}(b) shows \gls{squid}-oscillations centered at $\subtxt{I}{sw ,ref}$, coherent with an asymmetric \gls{squid} where the critical current can be expressed as \subtxt{I}{c, SQUID} $=$ \subtxt{I}{c, ref} + \subtxt{I}{c}$\sin{(\subtxt{\phi}{ext} + \subtxt{\varphi}{0})}$. 
\autoref{fig:fig3}(c) shows a similar plot at a different \subtxt{V}{TG}, with the oscillations still centered at $\subtxt{I}{} = \subtxt{I}{sw ,ref}$, but $\pi$ shifted with respect to \autoref{fig:fig3}(b). The extracted switching currents are plotted in \autoref{fig:fig3}(d) (dotted) together with sinusoidal fits (solid) yielding $\subtxt{I}{sw} = \SI{39}{\pico\ampere}$ (blue) and $\subtxt{I}{sw} = \SI{6.0}{\pico\ampere}$ (orange), respectively.\\
We study the sign of the \gls{qd}'s critical current in the gate range highlighted in grey in \autoref{fig:fig3}(a) by fixing the applied bias to $I = \SI{0.39}{\nano\ampere}$ and sweeping \subtxt{V}{TG} while \subtxt{\phi}{ext} is moved through zero. The color map in \autoref{fig:fig3}(e) indicates the voltage drop across the \gls{squid}, with blue (zero) marking the superconducting region, and white ($\neq$ zero) indicating the normal region. A change in the color implies a sign-change of \subtxt{I}{c} contributing to the total critical current \subtxt{I}{c,SQUID}. Comparing \autoref{fig:fig3}(a) and (e), \subtxt{V}{TG} not only changes the \gls{qd} occupation but also induces switches of the phase $\phi_0$ between $0$ and $\pi$. We note that the number of \glspl{cd} observed in \autoref{fig:fig3}(a) is half the number of phase shifts measured in the same gate range, indicating that the $\pi$ shift does not strictly depend on the parity of the ground state. Rather, the dynamics of the permutations and the symmetry of the orbitals involved in the transport across the \gls{qd} play an important role, as we will further discuss below.

\subsection{RF - SQUID}
\begin{figure}[ht!]
        \centering
        \includegraphics[width = 0.5\textwidth]{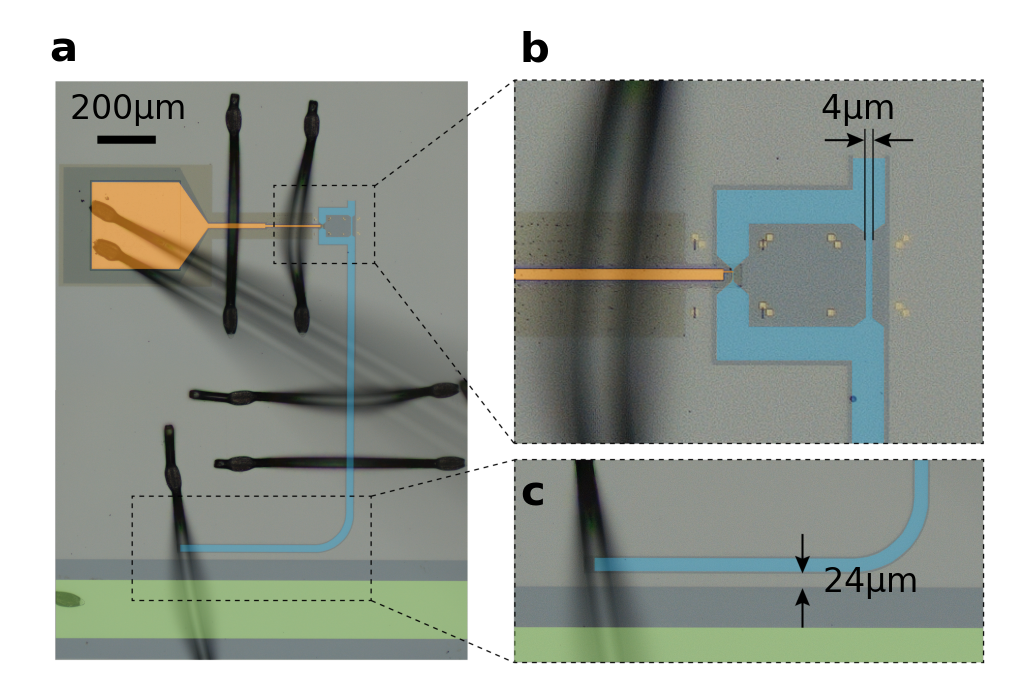}
        \caption{\textbf{\gls{rf} device}. (a) Optical image of the \gls{cptr} galvanically coupled to a \gls{rf}-\gls{squid}. Green, blue and orange indicate respectively the \gls{tl}, the \gls{cptr}-\gls{squid} and the gate metal. (b) Zoom-in of the \gls{squid} region. On the left side the \gls{jj}, on the right a constriction in the \gls{cptr} feedline acting as an inductor \subtxt{L}{ref}. (c) Zoom-in of the capacitive coupling between the \gls{tl} and the \gls{cptr} yielding a coupling quality factor \subtxt{Q}{c} $\simeq 10000$. }
        \label{fig:fig4m1}
\end{figure}
\begin{figure*}[ht!]
        \centering
        \includegraphics[width = 1.0\textwidth]{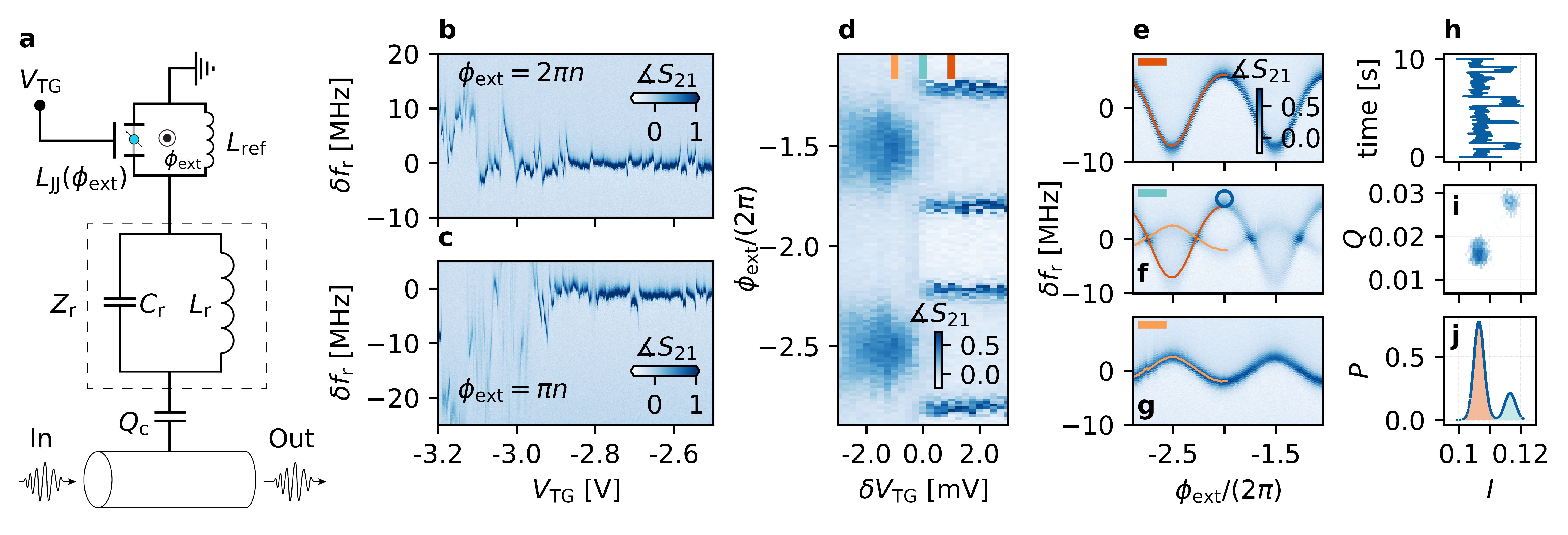}
        \caption{\textbf{0-$\pi$ transition in an \gls{rf}-\gls{squid}}. (a) Equivalent lumped elements circuit of the device shown in \autoref{fig:fig4m1}. (b) Spectrum as a function of \subtxt{V}{TG} at fixed flux \subtxt{\phi}{ext}$=2\pi n$. (c) Spectrum as a function of \subtxt{V}{TG} at fixed flux \subtxt{\phi}{ext}$=\pi n$. (d) Measured microwave phase $\measuredangle$\subtxt{S}{21} at fixed \subtxt{f}{p} as a function of \subtxt{V}{TG} and \subtxt{\phi}{ext}. The x-axis is centered around \subtxt{V}{TG} $=$\SI{-3.077}{\volt}. (e-g) Cuts marked in (d), solid lines are the extracted \subtxt{f}{r}. (e) and (g) are \gls{squid}-oscillations shifted by $\pi$ with respect to each other. (f) shows a cut at the transition, with the spectrum averaging across both states, the solid lines are a superposition of the \subtxt{f}{r} extracted in (e) and (g). (h) Filtered time trace of the real part $I$. The signals jumps frequently between the 0 and the $\pi$ phase state. (i) Time trace plotted in the $I$/$Q$ plane in blue and white for max and min counts, respectively.  (j) Normalized probability $P$ distribution extracted from the time trace.}
        \label{fig:fig4}
\end{figure*}
While the DC approach already gives interesting insights, clear limitations arise from the data acquisition rate and from the loss of information provided by measuring \subtxt{I}{sw} rather than directly probing \subtxt{I}{c}. Therefore we embed the \gls{jj} in an RF-\gls{squid} coupled to a superconducting \gls{cptr}, in order to probe the inductance of the \gls{jj}. We use a device as shown in \autoref{fig:fig4m1}, consisting of a $\lambda$/4 \gls{cptr} shorted to ground by an RF-\gls{squid}. The RF-\gls{squid} is composed of a gate-controlled \gls{jj} tuned by \subtxt{V}{TG}, and a thin \gls{al} stripe forming a fixed inductor. The inductance amount to \subtxt{L}{JJ}(\subtxt{\phi}{ext}) and $\subtxt{L}{ref} \simeq$\SI{1.5}{\nano\henry} for the \gls{jj} and the fixed reference inductor, respectively. The zoomed-in picture in \autoref{fig:fig4m1}(b) shows the RF-\gls{squid} and \autoref{fig:fig4m1}(c) displays a zoom-in of the area where the resonator is coupled to the transmission line. Our measurement signal consists of the transmitted signal \subtxt{S}{12} indicated in \autoref{fig:fig4}(a). A circle fit in the complex $I$/$Q$ plane yields the resonance frequency \subtxt{f}{r} of the whole device, the loaded quality factor \subtxt{Q}{l} and the average photon number \subtxt{n}{p}. The photon number is kept constant at approximately $\subtxt{n}{p} \simeq 5$ throughout the experiment.
In the following measurements, we systematically subtract \subtxt{f}{r,0}, the resonance frequency of the resonator at \subtxt{V}{TG}$= 0$, from the frequency of the probe signal \subtxt{f}{p} and from the resonance frequency of the \gls{rf}-device \subtxt{f}{r}.  \\
The resonance frequency \subtxt{f}{r} of the equivalent circuit in \autoref{fig:fig4}(a) can be expressed as:
\begin{equation}   
\label{eq:fr}
    \subtxt{f}{r}(\subtxt{\phi}{ext}) = \frac{1}{\sqrt{\Big(4l\subtxt{L}{r} + \frac{\pi^{2}}{2}\Big(\frac{1}{\subtxt{L}{JJ}(\subtxt{\phi}{ext}) + \subtxt{L}{l}} + \frac{1}{\subtxt{L}{ref}}\Big)^{-1}\Big)\subtxt{C}{r}4l}}
\end{equation}
where we have introduced the quantities $\subtxt{L}{l} \simeq$ \SI{0.5}{\nano\henry}, which denote the loop inductance of the \gls{squid} arm with the \gls{jj}. The total loop inductance \subtxt{L}{ref} + \subtxt{L}{l} yields $\simeq$ \SI{2.0}{\nano\henry}, with negligible effects considering the correction to the normalized external flux $\delta \subtxt{\phi}{ext} = - 2\pi(\subtxt{L}{ref}+\subtxt{L}{l})\subtxt{I}{c,max} \simeq 0.6\%$. Further, $l =$ \SI{1.538}{\milli\meter} is the length of the \gls{cptr} without the \gls{squid}, \subtxt{L}{r} $=$ \subtxt{L}{geo} + \subtxt{L}{kin} $\simeq$ \SI{1.724}{\micro\henry/\meter} and \subtxt{C}{r} $\simeq$ \SI{0.4}{\nano\farad/\meter} are respectively the inductance and capacitance of the \gls{cptr} per unit length. With that, the frequency response yields direct information on the \gls{jj} inductance.
The relation between \subtxt{L}{JJ}(\subtxt{\phi}{ext}) and the supercurrent \subtxt{I}{JJ}(\subtxt{\phi}{ext}) reads
\begin{equation}
\label{eq:Ljj}
    \frac{1}{\subtxt{L}{JJ}(\subtxt{\phi}{ext})} = \frac{2\pi}{\Phi_{\mathrm{0}}} \left. \frac{\partial \subtxt{I}{JJ}(x)}{\partial x}\right|_{\subtxt{\phi}{ext}},
\end{equation}
where \subtxt{I}{JJ}(\subtxt{\phi}{ext}) is given by the Fourier expansion:
\begin{equation}
    \label{eq:Ic}
    \subtxt{I}{JJ}(\subtxt{\phi}{ext}) = \sum_{k} (-1)^{k-1}\subtxt{A}{k}\sin{(k\subtxt{\phi}{ext} + \subtxt{\varphi}{0})}. 
\end{equation}
From \eqref{eq:Ic} and \eqref{eq:Ljj} we deduce that, at $\subtxt{\phi}{ext} = 2\pi n$, \subtxt{L}{JJ} has a local minimum or maximum for respectively a phase offset \subtxt{\varphi}{0} equal $0$ or $\pi$. In the resonance frequency, after subtracting \subtxt{f}{r,0}, this is visible as a sign change from negative to positive. At $\subtxt{\phi}{ext} = (2n+1)\pi$, the opposite effects are expected. \\
We plot the microwave phase response $\measuredangle$\subtxt{S}{21} in \autoref{fig:fig4}(b) to highlight the evolution of the resonance frequency (dark blue) as a function of applied \subtxt{V}{TG} at \subtxt{\phi}{ext} $= 2\pi n$. The resonance frequency tends to increase with the accumulation of charges at negative voltage, consistent with the increased supercurrent through the \gls{jj} and the accompanied decrease in \subtxt{L}{jj}. With the same reasoning, at \subtxt{\phi}{ext} $= (2n+1)\pi$ shown in \autoref{fig:fig4}(c), \subtxt{f}{r} tends to decrease. However, the resonance frequencies do not depend monotonically on \subtxt{V}{TG}, but they often cross zero before eventually staying above zero in \autoref{fig:fig4}(b) and below zero in \autoref{fig:fig4}(c). \\
We focus on the zero-crossing at \subtxt{V}{TG} = \SI{-3.077}{\volt} by fixing the probe frequency close to \subtxt{f}{r,0} and plotting the evolution of $\measuredangle$\subtxt{S}{21} as a function of \subtxt{V}{TG} and \subtxt{\phi}{ext} in \autoref{fig:fig4}(d). Around the zero-crossing of $\subtxt{\delta V}{TG} = \subtxt{V}{TG} + \SI{3.077}{\volt}$, we observe a qualitative change in the phase response of the junction, which we study in more detail by extracting \subtxt{f}{r}(\subtxt{\phi}{ext}). The cut in \autoref{fig:fig4}(e) at $\subtxt{\delta V}{TG} >0$ reveals oscillations of the supercurrent without phase offset, $\subtxt{\varphi}{0} = 0$. On the other hand, the cut $\subtxt{\delta V}{TG} <0$ in \autoref{fig:fig4}(g) shows $\pi$ shifted oscillations. The trace at $\subtxt{\delta V}{TG} =0$ shown in \autoref{fig:fig4}(f) reveals a coexistence of the signals shown in \autoref{fig:fig4}(e) and \autoref{fig:fig4}(g). In this situation, the zero and $\pi$ states are degenerate, and the microwave readout is sensitive enough to pick up the signals of both states. We fix the frequency of the probe at the circle indicated in \autoref{fig:fig4}(f), set the bandwidth to \SI{200}{\hertz} and measure the complex microwave response as a function of time. The real ($I$) and imaginary ($Q$) response reveal jumps between the 0 and $\pi$ phase state in \autoref{fig:fig4}(h), which are visible as two distinct regions in the $I$/$Q$ plane in \autoref{fig:fig4}(j), where blue indicates maximum number of counts. From the filtered data we extract the probability distribution $P$ of the two states in \autoref{fig:fig4}(i).

\subsection{$\pi$ - junction regimes}
\begin{figure}[ht!]
        \centering
        \includegraphics[width = 0.5\textwidth]{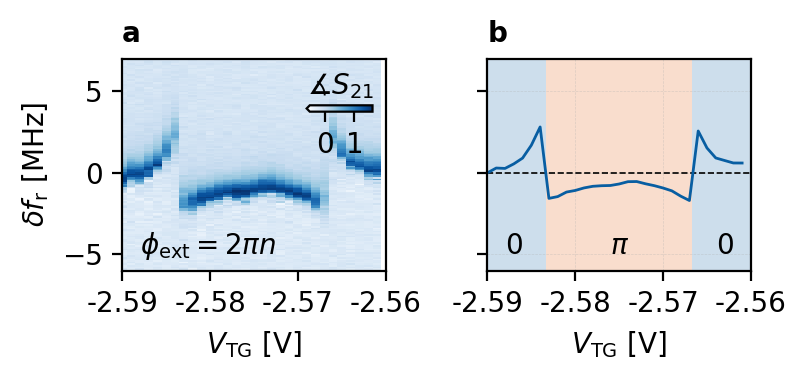}
        \caption{\textbf{Symmetry point}. (a) Spectrum as a function of gate voltage at fixed \subtxt{\phi}{ext} $= 2\pi n$. (b) Extracted \subtxt{f}{r} from (a). The sharp transition denotes a change in the \gls{qd} occupation and a consequent $0$ to $\pi$ transition in the \gls{jj}.} 
        \label{fig:fig6}
\end{figure}

\begin{figure*}[ht!]
        \centering
        \includegraphics[width = 1.0\textwidth]{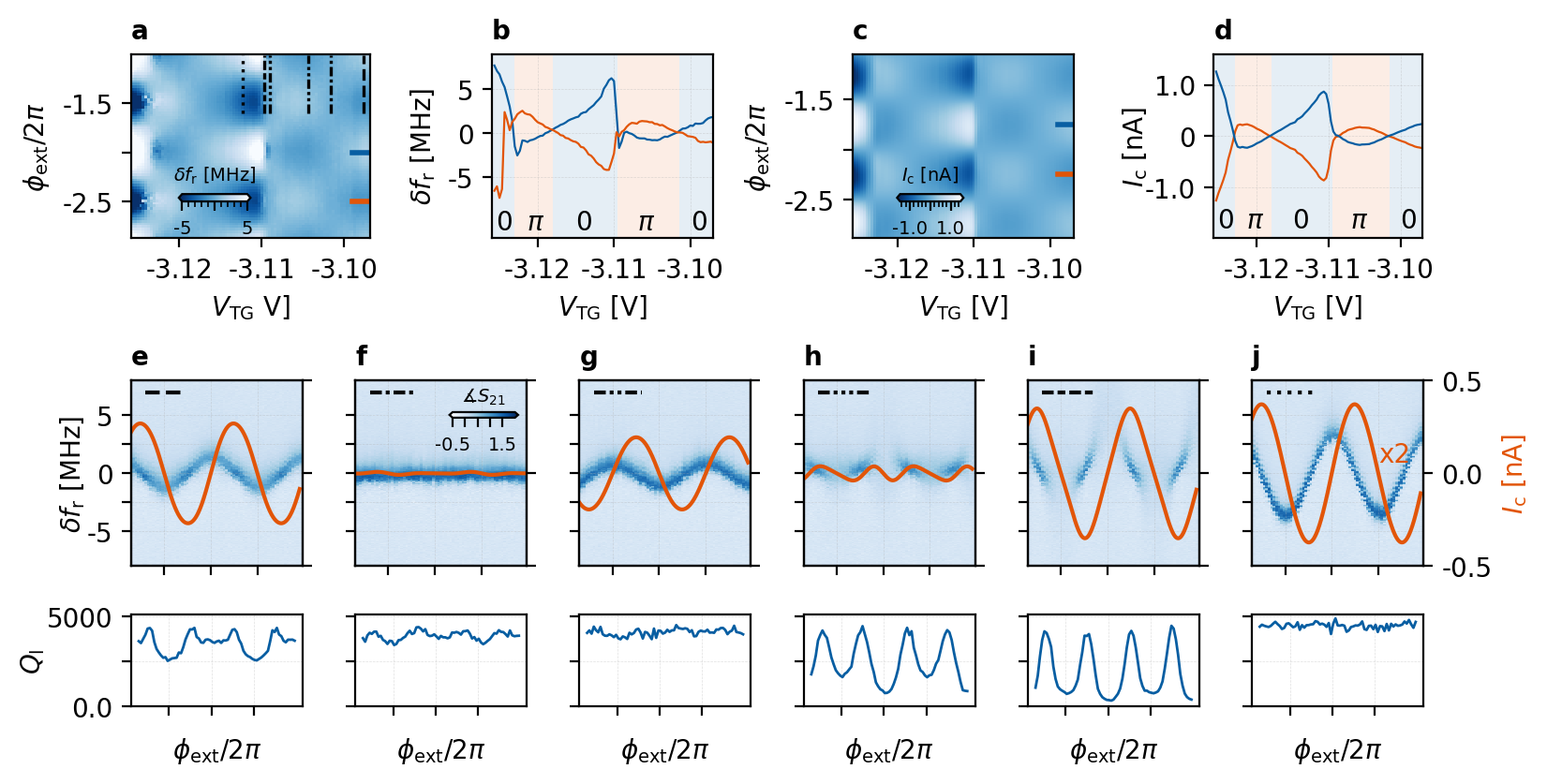}
        \caption{\textbf{Multi-level $\pi$-junction}. (a) \subtxt{f}{r} as a function of gate voltage and flux for a different range of voltage. (b) Cuts at fixed flux marked in (a). (c) Extracted \subtxt{I}{c} from the frequency response in (a). (d) Cuts at fixed flux marked in (c). (e-j) Spectrum as a function of flux of multiple cuts marked in (a). Bottom row shows the extracted loaded quality factor \subtxt{Q}{l} of the \gls{cptr}.} 
        \label{fig:fig7}
\end{figure*}
The transport properties of a \gls{sc} - \gls{qd} - \gls{sc} \gls{jj} strongly depend on the \gls{qd} single particle level spacing $\delta$ and its relation with the charging energy \subtxt{U}{} as sketched in \autoref{fig:HS}(b-d) \cite{shimizu_multilevel_1998}. In our device we tune all these parameters by varying \subtxt{V}{TG}. Based on our observations, we distinguish between two types of 0-$\pi$ transitions, namely sharp transitions and smooth transitions. A first example is shown in \autoref{fig:fig6}(a-b). \autoref{fig:fig6}(a) shows the frequency-dependent resonator response as a function of \subtxt{V}{TG} at fixed \subtxt{\phi}{ext}$=2\pi n$, while \autoref{fig:fig6}(b) shows the extracted resonance frequency. The regions of positive $\subtxt{f}{r} - \subtxt{f}{r,0}$ indicate a ground state with minimal energy at $\phi_0 = 0$, as argued above. The signal undergoes two sharp sign changes, indicating the formation of a $\pi$-junction in-between. \autoref{fig:fig6}(a-b) show an electron-hole symmetry point in which the $\pi$ shift is triggered by changing the occupation of the \gls{qd}.  \\
For a different regime of the device, we plot a map of the resonance frequency shift as a function of gate voltage and \subtxt{\phi}{ext} in \autoref{fig:fig7}(a). The corresponding critical current map is shown in \autoref{fig:fig7}(c). The frequency and the critical current clearly oscillate in \subtxt{\phi}{ext}, but the phase varies as a function of gate voltage. For example, we identify two sharp transitions at \subtxt{V}{TG} $\simeq$ \SI{-3.11}{\volt} and \SI{-3.123}{\volt}. However, between the sharp transitions, a smooth color change unveils an additional change in phase offset $\phi_0$. The frequency cuts in \autoref{fig:fig7}(b) at \subtxt{\phi}{ext} equal to zero (blue) and $\pi$  (orange) highlight both, the sharp and the smooth transitions between the ground states of the \gls{jj}. At the $0-\pi$ transitions, the supercurrent changes sign, as shown in \autoref{fig:fig7}(d). We plot in \autoref{fig:fig7}(e-j) in the upper row the flux dependence of the frequency shift at fixed gate values marked in \autoref{fig:fig7}(a). The bottom row shows the extracted loaded quality factor \subtxt{Q}{l} of the \gls{cptr} at the same gate voltage, respectively. \autoref{fig:fig7}(e-g) follow a smooth 0-$\pi$ transition and show that the amplitude of the \gls{squid}-oscillation goes to zero at the transition, see \autoref{fig:fig7}(f). On the other hand, \autoref{fig:fig7}(h-j) are taken across a sharp transition. Here, the oscillation amplitude decreases, but is not fully suppressed, as shown in \autoref{fig:fig7}(h). There, the coexistence of both states is visible by the switch of the \gls{jj} between 0 and $\pi$ as the flux changes \cite{delagrange_0_2018,sahu_ground-state_2024}. This observation indicates that the 0-$\pi$ transitions are qualitatively different. At the sharp transition, two ground states coexist, and the system switches between the two. At the smooth transition, the ground state remains the same, but the amplitude of different co-tunnelling contributions changes and averages to zero. A striking difference between smooth and sharp transitions is also found in the \subtxt{Q}{l} behavior as a function of flux. Considering the circuit in \autoref{fig:fig4}(a), the coexistence of two ground states would lead to a split of the resonance frequency and a consequent drop of \subtxt{Q}{l}, in the case of under coupled \gls{cptr} \cite{haller_phasedependent_2022}. However, at the smooth transition in \autoref{fig:fig7}(f), the quality factor does not drop significantly and, similar as with the frequency shift, the dependence of \subtxt{Q}{l} on flux is small. The dependence fully disappears away from the transition in \autoref{fig:fig7}(g) and reappears close to the sharp transition in \autoref{fig:fig7}(h). There, a strong modulation is observed in \autoref{fig:fig7}(i) with relative maximum at the flux matching the crossing of the 0 and $\pi$ states. Past the transition in \autoref{fig:fig7}(j), the dependence of \subtxt{Q}{l} on the flux vanishes again.

A similar phase response including sharp and smooth transitions is visible in the DC-device shown in \autoref{fig:fig3}(e). We compare the observed phase transitions with the position of the \glspl{cd} in \autoref{fig:fig3}(a) and find that the sharp transitions correspond to a change in occupation of the \gls{qd}. The smooth transitions occur at fixed number of charge, while \subtxt{V}{TG} still tunes the \gls{qd} energy. As explained in \autoref{fig:HS}, the alternation between sharp and smooth transitions is a fingerprint of a multi-level $\pi$-junction \cite{shimizu_multilevel_1998,vandam_supercurrent_2006,delagrange_0_2018}. In this type of junction, sharp transitions occur due to charges being loaded into the \gls{qd}, while smooth transitions arise when detuning the excited states taking part in the transport via co-tunneling, thereby changing the weights of participating co-tunnelling paths. We note that in this regime, also the parity of the involved orbitals can induce $\pi$ shifts \cite{shimizu_multilevel_1998}. 

\section{Discussion}
Using a combination of DC and RF experiments, we have drawn a clear picture of 0-$\pi$ transitions occurring in a \gls{sc} - \gls{qd} - \gls{sc} \gls{jj} in a planar heterostructure. In particular, we perform a detailed analysis of a multi-level $\pi$-junction, where 0-$\pi$-transitions not only occur due to a change in ground state parity, but also due to weight shifts of co-tunneling processes via low-energy excited \gls{qd} states. The fingerprints of the smooth transitions and their change in slope are inherently different from the better known $\pi$-junction with pronounced even-odd alternations. The latter occur in small \glspl{qd} where the single-particle level spacing is large, or in systems where different co-tunneling processes add incoherently and thus average out. In our experiment, we demonstrate the successful formation of a \gls{qd} in our hybrid heterostructure, in which tunneling paths through excited states add phase-coherently. 

Differences in the two types of transitions are visible also in the frequency shift and quality factor as a function of flux, insights that we obtained from the additional information encoded in the microwave signal. This demonstrates the potential of on-chip microwave technology for studying the properties of high quality \gls{qd} on planar \gls{ge}/\gls{sige} \glspl{qw}. Together with the large induced gap $\Delta^{*}$ of about \SI{240}{\micro\electronvolt}, our hybrid heterostructure is well suited for studying individual Andreev bound states, crossed Andreev reflection and eventually emerging topological states \cite{kitaev_unpaired_2001}.

\section{Methods}
\label{sec:fab}
All the measurements presented are performed in a \gls{dr} with a mixing chamber base temperature of \SI{70}{\milli\degreekelvin}. We use standard \gls{dc} measurements for the characterization of the DC-\gls{squid}. The RF-\gls{squid} transmission signal is measured using a Rohde\&Schwarz ZNB8 vector network analyzer. The input signal is first attenuated at each stage of the \gls{dr} while the output is amplified using a Low Noise Factory LNF\_LNC1\_12A \gls{hemt} amplifier, the signal is amplified again at room temperature with an additional amplifier Miteq AMF-3F-01000400-08-10P.  The external flux \subtxt{\phi}{ext} is supplied by a homemade superconducting coil mounted on the back of the \gls{pcb}. An \gls{al} magnetic field shield encapsulates all the components. See Supplementary for a sketch of the whole setup and all the details. 
Fabrication is carried out using standard E-beam/photo - lithography technique. The \gls{al} is etched using standard Transene-D wet-etching. The \gls{hs} mesa is \gls{rie} etched using \subtxt{\mathrm{SF}}{6}/\subtxt{\mathrm{O}}{2} chemical. Gate stack consists of a 300 cycles \gls{ald} \subtxt{\mathrm{Al}}{2}\subtxt{\mathrm{O}}{3} deposited at \SI{90}{\degree} and \SI{100}{\nano\meter}(DC device) and \SI{150}{\nano\meter}(\gls{rf} device) E-beam evaporated \gls{al}. 

\section*{Acknowledgements}
This work was supported as a part of NCCR SPIN, a National Centre of Competence in Research, funded by the Swiss National Science Foundation (grant number 225153), by the European Union’s Horizon 2020 research and innovation program through the Marie Sk\l{}odowska Curie COFUND project QUSTEC, agreement No 847471, by the Swiss National Science Foundation through grant 192027 and by the Basel QCQT PhD school. The research is part of the Munich Quantum Valley, which is supported by the Bavarian state government with funds from the Hightech Agenda Bavaria.

\bibliography{Pi_bibliography}

@article{jirovec_singlettriplet_2021b,
	title = {A singlet-triplet hole spin qubit in planar {Ge}},
	volume = {20},
	url = {https://www.nature.com/articles/s41563-021-01022-2},
	doi = {10.1038/s41563-021-01022-2},
	number = {8},
	journal = {Nature Materials},
	author = {Jirovec, Daniel and Hofmann, Andrea and Ballabio, Andrea and Mutter, Philipp M. and Tavani, Giulio and Botifoll, Marc and Crippa, Alessandro and Kukucka, Josip and Sagi, Oliver and Martins, Frederico and Saez-Mollejo, Jaime and Prieto, Ivan and Borovkov, Maksim and Arbiol, Jordi and Chrastina, Daniel and Isella, Giovanni and Katsaros, Georgios},
	month = aug,
	year = {2021},
	pages = {1106--1112},
}

@article{nigro_demonstration_2024b,
	title = {Demonstration of {Microwave} {Resonators} and {Double} {Quantum} {Dots} on {Optimized} {Reverse}-{Graded} {Ge}/{SiGe} {Heterostructures}},
	volume = {6},
	url = {https://doi.org/10.1021/acsaelm.4c00654},
	doi = {10.1021/acsaelm.4c00654},
	number = {7},
	urldate = {2025-06-10},
	journal = {ACS Applied Electronic Materials},
	author = {Nigro, Arianna and Jutzi, Eric and Oppliger, Fabian and De Palma, Franco and Olsen, Christian and Ruiz-Caridad, Alicia and Gadea, Gerard and Scarlino, Pasquale and Zardo, Ilaria and Hofmann, Andrea},
	month = jul,
	year = {2024},
	pages = {5094--5100},
}

@article{kiyooka_gatemon_2025a,
	title = {Gatemon {Qubit} on a {Germanium} {Quantum}-{Well} {Heterostructure}},
	volume = {25},
	issn = {1530-6984},
	url = {https://doi.org/10.1021/acs.nanolett.4c05539},
	doi = {10.1021/acs.nanolett.4c05539},
	number = {1},
	urldate = {2025-07-29},
	journal = {Nano Letters},
	author = {Kiyooka, Elyjah and Tangchingchai, Chotivut and Noirot, Leo and Leblanc, Axel and Brun, Boris and Zihlmann, Simon and Maurand, Romain and Schmitt, Vivien and Dumur, Étienne and Hartmann, Jean-Michel and Lefloch, Francois and De Franceschi, Silvano},
	month = jan,
	year = {2025},
	pages = {562--568},
}

@article{defranceschi_hybrid_2010,
  title = {Hybrid Superconductor--Quantum Dot Devices},
  author = {De Franceschi, Silvano and Kouwenhoven, Leo and Sch{\"o}nenberger, Christian and Wernsdorfer, Wolfgang},
  year = 2010,
  month = oct,
  journal = {Nature Nanotechnology},
  volume = {5},
  number = {10},
  pages = {703--711},
  doi = {10.1038/nnano.2010.173}
}

@article{cleuziou_carbon_2006,
  title = {Carbon Nanotube Superconducting Quantum Interference Device},
  author = {Cleuziou, J.-P. and Wernsdorfer, W. and Bouchiat, V. and Ondar{\c c}uhu, T. and Monthioux, M.},
  year = 2006,
  month = oct,
  journal = {Nature Nanotechnology},
  volume = {1},
  number = {1},
  pages = {53--59},
  issn = {1748-3387, 1748-3395},
  doi = {10.1038/nnano.2006.54},
  urldate = {2026-06-17},
  copyright = {http://www.springer.com/tdm},
  langid = {english}
}

@article{lee_spinresolved_2014a,
  title = {Spin-Resolved {{Andreev}} Levels and Parity Crossings in Hybrid Superconductor--Semiconductor Nanostructures},
  author = {Lee, Eduardo J. H. and Jiang, Xiaocheng and Houzet, Manuel and Aguado, Ram{\'o}n and Lieber, Charles M. and De Franceschi, Silvano},
  year = 2014,
  month = jan,
  journal = {Nature Nanotechnology},
  volume = {9},
  number = {1},
  pages = {79--84},
  issn = {1748-3387, 1748-3395},
  doi = {10.1038/nnano.2013.267},
  urldate = {2026-06-17},
  copyright = {http://www.springer.com/tdm},
  langid = {english}
}

@article{pillet_andreev_2010,
  title = {Andreev Bound States in Supercurrent-Carrying Carbon Nanotubes Revealed},
  author = {Pillet, J-D. and Quay, C. H. L. and Morfin, P. and Bena, C. and Yeyati, A. Levy and Joyez, P.},
  year = 2010,
  month = dec,
  journal = {Nature Physics},
  volume = {6},
  number = {12},
  pages = {965--969},
  doi = {10.1038/nphys1811}
}

@article{eichler_evenodd_2007,
  title = {Even-{{Odd Effect}} in {{Andreev Transport}} through a {{Carbon Nanotube Quantum Dot}}},
  author = {Eichler, A. and Weiss, M. and Oberholzer, S. and Sch{\"o}nenberger, C. and Levy Yeyati, A. and Cuevas, J. C. and {Mart{\'i}n-Rodero}, A.},
  year = 2007,
  month = sep,
  journal = {Physical Review Letters},
  volume = {99},
  number = {12},
  pages = {126602},
  issn = {0031-9007, 1079-7114},
  doi = {10.1103/PhysRevLett.99.126602},
  urldate = {2026-06-17},
  copyright = {http://link.aps.org/licenses/aps-default-license},
  langid = {english}
}

@article{vanheck_conductance_2016,
  title = {Conductance of a Proximitized Nanowire in the {{Coulomb}} Blockade Regime},
  author = {Van Heck, B. and Lutchyn, R. M. and Glazman, L. I.},
  year = 2016,
  month = jun,
  journal = {Physical Review B},
  volume = {93},
  number = {23},
  pages = {235431},
  issn = {2469-9950, 2469-9969},
  doi = {10.1103/PhysRevB.93.235431},
  urldate = {2026-06-17},
  copyright = {http://link.aps.org/licenses/aps-default-license},
  langid = {english}
}

@article{echternach_progress_2009,
  title = {Progress in the Development of a Single {{Cooper-pair}} Box Qubit},
  author = {Echternach, Pierre M. and Schneiderman, J. F. and Shaw, Matthew D. and Delsing, Per},
  year = 2009,
  month = jun,
  journal = {Quantum Information Processing},
  volume = {8},
  number = {2-3},
  pages = {183--198},
  issn = {1570-0755, 1573-1332},
  doi = {10.1007/s11128-009-0097-x},
  urldate = {2026-06-17},
  copyright = {http://www.springer.com/tdm},
  langid = {english}
}

@misc{fabris_granular_2026,
  title = {Granular Aluminum Induced Superconductivity in Germanium for Hole Spin-Based Hybrid Devices},
  author = {Fabris, Giorgio and {Falthansl-Scheinecker}, Paul and Shah, Devashish and Pino, Daniel Michel and Borovkov, Maksim and Bubis, Anton and Roux, Kevin and Sokolova, Dina and Juanes, Alejandro Andres and Costanzo, Tommaso and Taha, Inas and Gen{\c c}, Aziz and Arbiol, Jordi and Calcaterra, Stefano and Oliveira, Afonso De Cerdeira and Chrastina, Daniel and Isella, Giovanni and Souto, Ruben Seoane and Calderon, Maria Jose and Aguado, Ramon and {Abadillo-Uriel}, Jose Carlos and Katsaros, Georgios},
  year = 2026,
  publisher = {arXiv},
  doi = {10.48550/ARXIV.2602.21364},
  urldate = {2026-06-17},
  copyright = {Creative Commons Attribution 4.0 International}
}

@article{hendrickx_fourqubit_2021b,
	title = {A four-qubit germanium quantum processor},
	volume = {591},
	copyright = {2021 The Author(s), under exclusive licence to Springer Nature Limited},
	issn = {1476-4687},
	url = {https://www.nature.com/articles/s41586-021-03332-6},
	doi = {10.1038/s41586-021-03332-6},
	language = {en},
	number = {7851},
	urldate = {2023-08-28},
	journal = {Nature},
	author = {Hendrickx, Nico W. and Lawrie, William I. L. and Russ, Maximilian and van Riggelen, Floor and de Snoo, Sander L. and Schouten, Raymond N. and Sammak, Amir and Scappucci, Giordano and Veldhorst, Menno},
	month = mar,
	year = {2021},
	pages = {580--585},
}

@article{hinderling_direct_2024a,
	title = {Direct {Microwave} {Spectroscopy} of {Andreev} {Bound} {States} in {Planar} {Ge} {Josephson} {Junctions}},
	volume = {5},
	issn = {2691-3399},
	url = {https://link.aps.org/doi/10.1103/PRXQuantum.5.030357},
	doi = {10.1103/PRXQuantum.5.030357},
	language = {en},
	number = {3},
	urldate = {2025-12-02},
	journal = {PRX Quantum},
	author = {Hinderling, M. and Ten Kate, S. C. and Coraiola, M. and Haxell, D.Z. and Stiefel, M. and Mergenthaler, M. and Paredes, S. and Bedell, S.W. and Sabonis, D. and Nichele, F.},
	month = sep,
	year = {2024},
	pages = {030357},
}

@article{hendrickx_gatecontrolled_2018,
	title = {Gate-controlled quantum dots and superconductivity in planar germanium},
	volume = {9},
	issn = {2041-1723},
	url = {https://www.nature.com/articles/s41467-018-05299-x},
	doi = {10.1038/s41467-018-05299-x},
	language = {en},
	number = {1},
	urldate = {2025-12-02},
	journal = {Nature Communications},
	author = {Hendrickx, N. W. and Franke, D. P. and Sammak, A. and Kouwenhoven, M. and Sabbagh, D. and Yeoh, L. and Li, R. and Tagliaferri, M. L. V. and Virgilio, M. and Capellini, G. and Scappucci, G. and Veldhorst, M.},
	month = jul,
	year = {2018},
	pages = {2835},
}

@article{vandam_supercurrent_2006,
	title = {Supercurrent reversal in quantum dots},
	volume = {442},
	copyright = {http://www.springer.com/tdm},
	issn = {0028-0836, 1476-4687},
	url = {https://www.nature.com/articles/nature05018},
	doi = {10.1038/nature05018},
	language = {en},
	number = {7103},
	urldate = {2025-12-02},
	journal = {Nature},
	author = {Van Dam, Jorden A. and Nazarov, Yuli V. and Bakkers, Erik P. A. M. and De Franceschi, Silvano and Kouwenhoven, Leo P.},
	month = aug,
	year = {2006},
	pages = {667--670},
}

@article{sagi_gate_2024b,
	title = {A gate tunable transmon qubit in planar {Ge}},
	volume = {15},
	issn = {2041-1723},
	url = {https://www.nature.com/articles/s41467-024-50763-6},
	doi = {10.1038/s41467-024-50763-6},
	language = {en},
	number = {1},
	urldate = {2025-12-02},
	journal = {Nature Communications},
	author = {Sagi, Oliver and Crippa, Alessandro and Valentini, Marco and Janik, Marian and Baghumyan, Levon and Fabris, Giorgio and Kapoor, Lucky and Hassani, Farid and Fink, Johannes and Calcaterra, Stefano and Chrastina, Daniel and Isella, Giovanni and Katsaros, Georgios},
	month = jul,
	year = {2024},
	pages = {6400},
}

@article{vigneau_germanium_2019b,
	title = {Germanium {Quantum}-{Well} {Josephson} {Field}-{Effect} {Transistors} and {Interferometers}},
	volume = {19},
	copyright = {https://doi.org/10.15223/policy-029},
	issn = {1530-6984, 1530-6992},
	url = {https://pubs.acs.org/doi/10.1021/acs.nanolett.8b04275},
	doi = {10.1021/acs.nanolett.8b04275},
	language = {en},
	number = {2},
	urldate = {2025-12-02},
	journal = {Nano Letters},
	author = {Vigneau, Florian and Mizokuchi, Raisei and Zanuz, Dante Colao and Huang, Xuhai and Tan, Susheng and Maurand, Romain and Frolov, Sergey and Sammak, Amir and Scappucci, Giordano and Lefloch, Francois and De Franceschi, Silvano},
	month = feb,
	year = {2019},
	pages = {1023--1027},
}

@article{bargerbos_spectroscopy_2023,
	title = {Spectroscopy of {Spin}-{Split} {Andreev} {Levels} in a {Quantum} {Dot} with {Superconducting} {Leads}},
	volume = {131},
	issn = {0031-9007, 1079-7114},
	url = {https://link.aps.org/doi/10.1103/PhysRevLett.131.097001},
	doi = {10.1103/PhysRevLett.131.097001},
	language = {en},
	number = {9},
	urldate = {2025-12-02},
	journal = {Physical Review Letters},
	author = {Bargerbos, Arno and Pita-Vidal, Marta and Žitko, Rok and Splitthoff, Lukas J. and Grünhaupt, Lukas and Wesdorp, Jaap J. and Liu, Yu and Kouwenhoven, Leo P. and Aguado, Ramón and Andersen, Christian Kraglund and Kou, Angela and Van Heck, Bernard},
	month = aug,
	year = {2023},
	pages = {097001},
}

@misc{kate_finite_2025,
	title = {Finite {Length} {Effects} and {Coulomb} {Interaction} in {Ge} {Quantum} {Well}-{Based} {Josephson} {Junctions} {Probed} with {Microwave} {Spectroscopy}},
	url = {http://arxiv.org/abs/2508.06180},
	doi = {10.48550/arXiv.2508.06180},
	language = {en},
	urldate = {2025-12-02},
	publisher = {arXiv},
	author = {Kate, S. C. ten and Ohnmacht, D. C. and Coraiola, M. and Antonelli, T. and Paredes, S. and Schupp, F. J. and Hinderling, M. and Bedell, S. W. and Belzig, W. and Cuevas, J. C. and Svetogorov, A. E. and Nichele, F. and Sabonis, D.},
	month = aug,
	year = {2025},
	note = {arXiv:2508.06180 [cond-mat]},
}

@article{aggarwal_enhancement_2021b,
	title = {Enhancement of proximity-induced superconductivity in a planar {Ge} hole gas},
	volume = {3},
	issn = {2643-1564},
	url = {https://link.aps.org/doi/10.1103/PhysRevResearch.3.L022005},
	doi = {10.1103/PhysRevResearch.3.L022005},
	language = {en},
	number = {2},
	urldate = {2025-12-02},
	journal = {Physical Review Research},
	author = {Aggarwal, Kushagra and Hofmann, Andrea and Jirovec, Daniel and Prieto, Ivan and Sammak, Amir and Botifoll, Marc and Martí-Sánchez, Sara and Veldhorst, Menno and Arbiol, Jordi and Scappucci, Giordano and Danon, Jeroen and Katsaros, Georgios},
	month = apr,
	year = {2021},
	pages = {L022005},
}

@article{valentini_parityconserving_2024a,
	title = {Parity-conserving {Cooper}-pair transport and ideal superconducting diode in planar germanium},
	volume = {15},
	issn = {2041-1723},
	url = {https://www.nature.com/articles/s41467-023-44114-0},
	doi = {10.1038/s41467-023-44114-0},
	language = {en},
	number = {1},
	urldate = {2025-12-15},
	journal = {Nature Communications},
	author = {Valentini, Marco and Sagi, Oliver and Baghumyan, Levon and De Gijsel, Thijs and Jung, Jason and Calcaterra, Stefano and Ballabio, Andrea and Aguilera Servin, Juan and Aggarwal, Kushagra and Janik, Marian and Adletzberger, Thomas and Seoane Souto, Rubén and Leijnse, Martin and Danon, Jeroen and Schrade, Constantin and Bakkers, Erik and Chrastina, Daniel and Isella, Giovanni and Katsaros, Georgios},
	month = jan,
	year = {2024},
	pages = {169},
}

@article{lakic_quantum_2025,
	title = {A quantum dot in germanium proximitized by a superconductor},
	volume = {24},
	issn = {1476-1122, 1476-4660},
	url = {https://www.nature.com/articles/s41563-024-02095-5},
	doi = {10.1038/s41563-024-02095-5},
	language = {en},
	number = {4},
	urldate = {2025-12-15},
	journal = {Nature Materials},
	author = {Lakic, Lazar and Lawrie, William I. L. and Van Driel, David and Stehouwer, Lucas E. A. and Su, Yao and Veldhorst, Menno and Scappucci, Giordano and Kuemmeth, Ferdinand and Chatterjee, Anasua},
	month = apr,
	year = {2025},
	pages = {552--558},
}

@misc{hofmann_assessing_2019a,
	title = {Assessing the potential of {Ge}/{SiGe} quantum dots as hosts for singlet-triplet qubits},
	url = {http://arxiv.org/abs/1910.05841},
	doi = {10.48550/arXiv.1910.05841},
	language = {en},
	urldate = {2026-02-03},
	publisher = {arXiv},
	author = {Hofmann, Andrea and Jirovec, Daniel and Borovkov, Maxim and Prieto, Ivan and Ballabio, Andrea and Frigerio, Jacopo and Chrastina, Daniel and Isella, Giovanni and Katsaros, Georgios},
	month = oct,
	year = {2019},
	note = {arXiv:1910.05841 [cond-mat]},
}

@article{hinderling_flipchipbased_2024,
	title = {Flip-{Chip}-{Based} {Fast} {Inductive} {Parity} {Readout} of a {Planar} {Superconducting} {Island}},
	volume = {5},
	issn = {2691-3399},
	url = {https://link.aps.org/doi/10.1103/PRXQuantum.5.030337},
	doi = {10.1103/PRXQuantum.5.030337},
	language = {en},
	number = {3},
	urldate = {2026-03-27},
	journal = {PRX Quantum},
	author = {Hinderling, M. and Kate, S.C. Ten and Haxell, D.Z. and Coraiola, M. and Paredes, S. and Cheah, E. and Krizek, F. and Schott, R. and Wegscheider, W. and Sabonis, D. and Nichele, F.},
	month = aug,
	year = {2024},
	pages = {030337},
}

@article{shimizu_multilevel_1998,
	title = {Multilevel {Effect} on the {Josephson} {Current} through a {Quantum} {Dot}},
	volume = {67},
	issn = {0031-9015, 1347-4073},
	url = {http://journals.jps.jp/doi/10.1143/JPSJ.67.1525},
	doi = {10.1143/JPSJ.67.1525},
	language = {en},
	number = {5},
	urldate = {2026-04-24},
	journal = {Journal of the Physical Society of Japan},
	author = {Shimizu, Yoshiteru and Horii, Hideto and Takane, Yositake and Isawa, Yoshimasa},
	month = may,
	year = {1998},
	pages = {1525--1528},
}

@article{debbarma_josephson_2023,
	title = {Josephson {Junction} π − 0 {Transition} {Induced} by {Orbital} {Hybridization} in a {Double} {Quantum} {Dot}},
	volume = {131},
	issn = {0031-9007, 1079-7114},
	url = {https://link.aps.org/doi/10.1103/PhysRevLett.131.256001},
	doi = {10.1103/PhysRevLett.131.256001},
	language = {en},
	number = {25},
	urldate = {2026-04-24},
	journal = {Physical Review Letters},
	author = {Debbarma, Rousan and Tsintzis, Athanasios and Aspegren, Markus and Souto, Rubén Seoane and Lehmann, Sebastian and Dick, Kimberly and Leijnse, Martin and Thelander, Claes},
	month = dec,
	year = {2023},
	pages = {256001},
}

@article{haller_phasedependent_2022,
	title = {Phase-dependent microwave response of a graphene {Josephson} junction},
	volume = {4},
	issn = {2643-1564},
	url = {https://link.aps.org/doi/10.1103/PhysRevResearch.4.013198},
	doi = {10.1103/PhysRevResearch.4.013198},
	language = {en},
	number = {1},
	urldate = {2026-04-24},
	journal = {Physical Review Research},
	author = {Haller, R. and Fülöp, G. and Indolese, D. and Ridderbos, J. and Kraft, R. and Cheung, L. Y. and Ungerer, J. H. and Watanabe, K. and Taniguchi, T. and Beckmann, D. and Danneau, R. and Virtanen, P. and Schönenberger, C.},
	month = mar,
	year = {2022},
	pages = {013198},
}

@article{delagrange_0_2018,
	title = {0 − π {Quantum} transition in a carbon nanotube {Josephson} junction: {Universal} phase dependence and orbital degeneracy},
	volume = {536},
	issn = {09214526},
	shorttitle = {0 − π {Quantum} transition in a carbon nanotube {Josephson} junction},
	url = {https://linkinghub.elsevier.com/retrieve/pii/S0921452617306208},
	doi = {10.1016/j.physb.2017.09.034},
	language = {en},
	urldate = {2026-04-24},
	journal = {Physica B: Condensed Matter},
	author = {Delagrange, R. and Weil, R. and Kasumov, A. and Ferrier, M. and Bouchiat, H. and Deblock, R.},
	month = may,
	year = {2018},
	pages = {211--222},
}

@article{mannila_detecting_2019,
	title = {Detecting parity effect in a superconducting device in the presence of parity switches},
	volume = {100},
	issn = {2469-9950, 2469-9969},
	url = {https://link.aps.org/doi/10.1103/PhysRevB.100.020502},
	doi = {10.1103/PhysRevB.100.020502},
	language = {en},
	number = {2},
	urldate = {2026-04-25},
	journal = {Physical Review B},
	author = {Mannila, E. T. and Maisi, V. F. and Nguyen, H. Q. and Marcus, C. M. and Pekola, J. P.},
	month = jul,
	year = {2019},
	pages = {020502},
}

@article{aumentado_nonequilibrium_2004,
	title = {Nonequilibrium {Quasiparticles} and 2 e {Periodicity} in {Single}-{Cooper}-{Pair} {Transistors}},
	volume = {92},
	copyright = {http://link.aps.org/licenses/aps-default-license},
	issn = {0031-9007, 1079-7114},
	url = {https://link.aps.org/doi/10.1103/PhysRevLett.92.066802},
	doi = {10.1103/PhysRevLett.92.066802},
	language = {en},
	number = {6},
	urldate = {2026-04-25},
	journal = {Physical Review Letters},
	author = {Aumentado, J. and Keller, Mark W. and Martinis, John M. and Devoret, M. H.},
	month = feb,
	year = {2004},
	pages = {066802},
}

@article{sahu_ground-state_2024,
	title = {Ground-state phase diagram and parity-flipping microwave transitions in a gate-tunable {Josephson} junction},
	volume = {109},
	issn = {2469-9950, 2469-9969},
	url = {https://link.aps.org/doi/10.1103/PhysRevB.109.134506},
	doi = {10.1103/PhysRevB.109.134506},
	language = {en},
	number = {13},
	urldate = {2026-04-25},
	journal = {Physical Review B},
	author = {Sahu, M. R. and Matute-Cañadas, F. J. and Benito, M. and Krogstrup, P. and Nygård, J. and Goffman, M. F. and Urbina, C. and Yeyati, A. Levy and Pothier, H.},
	month = apr,
	year = {2024},
	pages = {134506},
}

@article{kitaev_unpaired_2001,
	title = {Unpaired {Majorana} fermions in quantum wires},
	volume = {44},
	issn = {1468-4780},
	url = {https://www.mathnet.ru/eng/ufn5648},
	doi = {10.1070/1063-7869/44/10S/S29},
	number = {10S},
	urldate = {2026-04-27},
	journal = {Physics-Uspekhi},
	author = {Kitaev, A Yu},
	month = oct,
	year = {2001},
	pages = {131--136},
}

@misc{ruggiero_high-quality_2025,
	title = {High-quality and field resilient microwave resonators on {Ge}/{SiGe} quantum well heterostructures},
	url = {http://arxiv.org/abs/2512.20238},
	doi = {10.48550/arXiv.2512.20238},
	urldate = {2026-04-27},
	publisher = {arXiv},
	author = {Ruggiero, Luigi and Ciacca, Carlo and Drexler, Pauline and Weibel, Vera Jo and Olsen, Christian and Schönenberger, Christian and Bougeard, Dominique and Hofmann, Andrea},
	month = dec,
	year = {2025},
}

\end{document}


\clearpage 
\thispagestyle{empty} 
\phantom{} 
\clearpage
\title{Supplementary Information}

\author{Luigi Ruggiero\textsuperscript{1}}
\author{Vera Jo Weibel\textsuperscript{1}}
\author{Pauline Drexler\textsuperscript{2}}
\author{Carlo Ciaccia\textsuperscript{1}}
\author{Christian Olsen\textsuperscript{1}}
\author{Dominique Bougeard\textsuperscript{2}}
\author{Christian Schönenberger\textsuperscript{3, 4}}
\author{Andrea Hofmann\textsuperscript{1, 3, *}}

\begin{abstract} 
\begin{center}
\textsuperscript{1}  University of Basel, Klingelbergstrasse 82, 4056 Basel, Switzerland \\
\textsuperscript{2} University of Regensburg, Universitätsstraße 31, 93053 Regensburg, Germany \\
\textsuperscript{3} Swiss Nanoscience Institute, Klingelbergstrasse 82, 4056 Basel, Switzerland\\
\textsuperscript{4} YQuantum, Parkstrasse 1, 5234 Villigen, Switzerland\\
\textsuperscript{*} Corresponding author. Email: andrea.hofmann@unibas.ch\\
\vspace{1em}

\date{\today}
\end{center}
\end{abstract}


\maketitle 

\subsection{Supplementary Figure 1: Junction behaviour in an extended gate range}
\begin{figure*}[ht!]
        \centering
        \includegraphics[width = 1.0\textwidth]{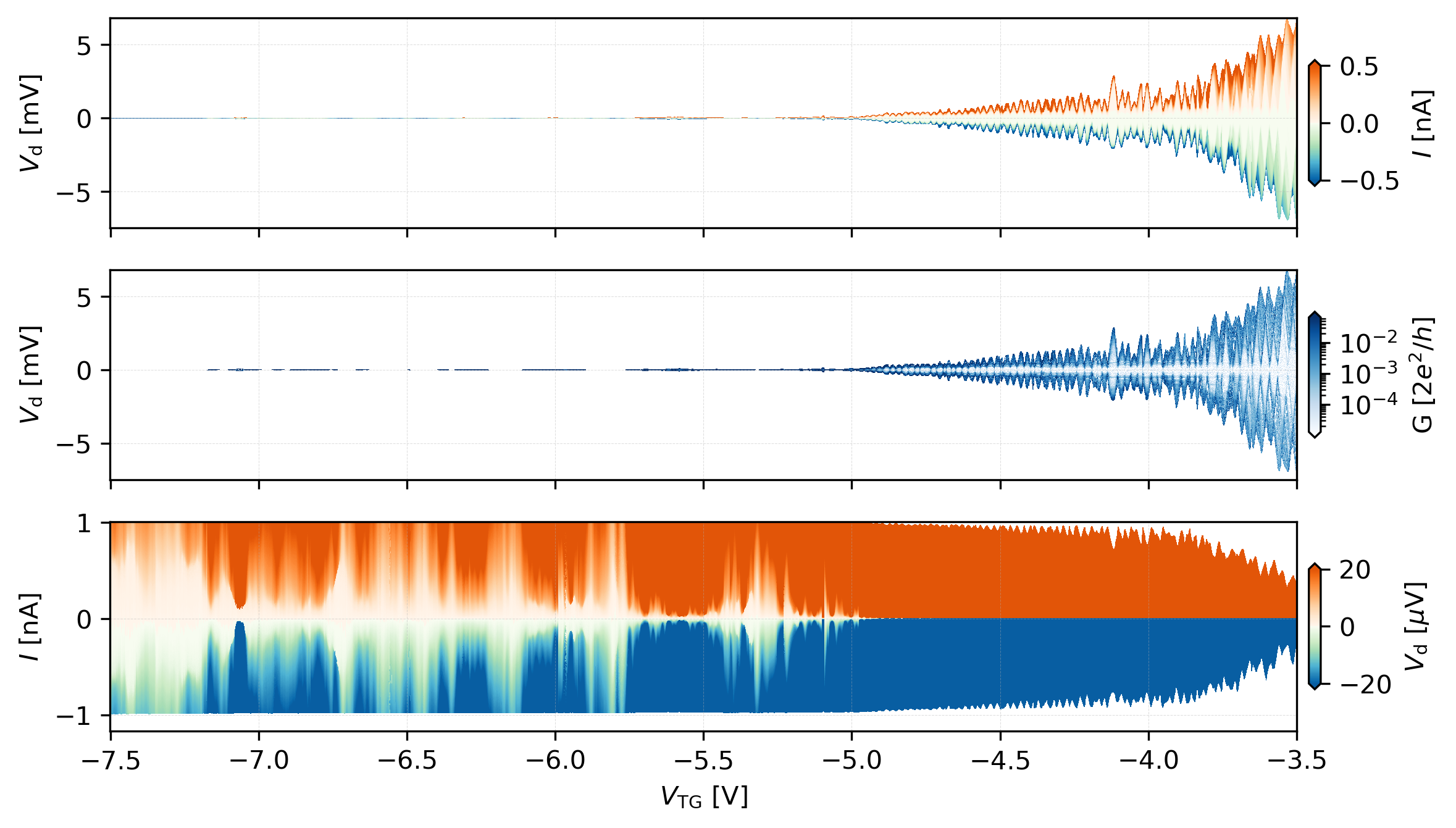}
        \caption{\textbf{From \gls{qd} to \gls{jj}}. (a) Measured bias current as a function of applied gate voltage \subtxt{V}{TG} and measured voltage drop \subtxt{V}{d} accross the \gls{jj}. (b) Same range of \subtxt{V}{TG} as in (a) where the colorbar refers to the differential conductance $G$. (c) Same as in (a) but with \textit{y} and \textit{z} axis flipped. The color bar limits have been reduced to improve visibility of the superconducting region, indicated as white around zero current $I$.}
        \label{fig:QD_to_JJ}
\end{figure*}

\newpage
\subsection{Supplementary Figure 2: Setup}
\begin{figure*}[ht!]
        \centering
        \includegraphics[width = 1.0\textwidth]{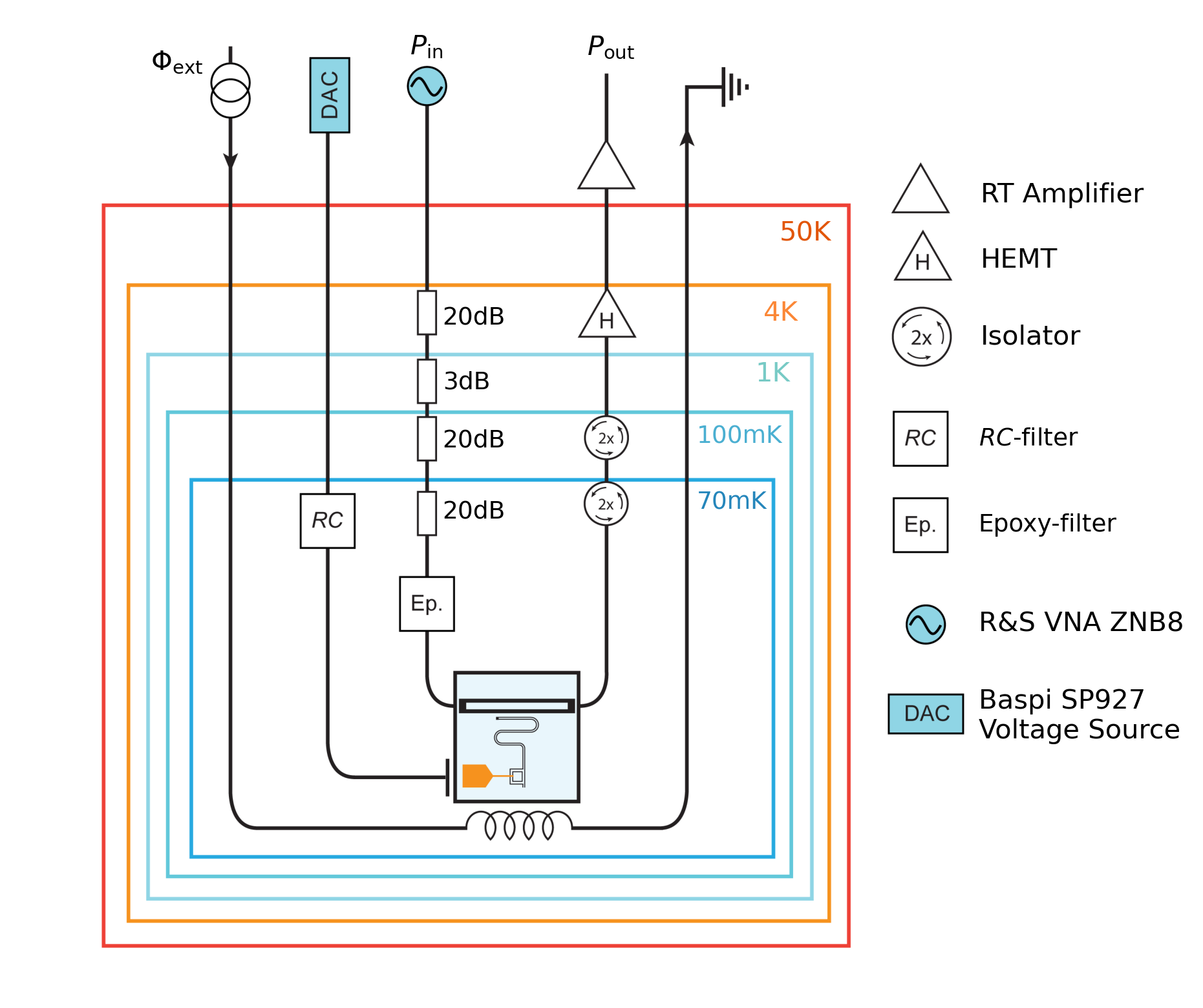}
        \caption{\textbf{Microwave measurement setup}. Complete setup for microwave characterization.}
        \label{fig:setup}
\end{figure*}

\newpage
\subsection{Supplementary Figure 3: Curve Fit}
\begin{figure*}[ht!]
        \centering
        \includegraphics[width = 1.0\textwidth]{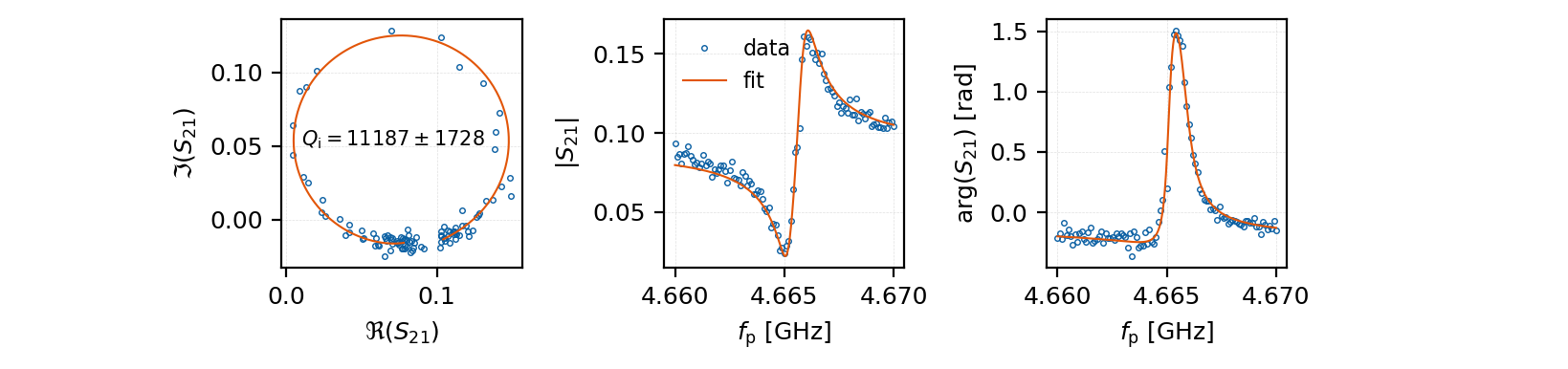}
        \caption{\textbf{Fitting routine on \subtxt{S}{21}}. (a) Circle fit of the resonance in the $I$(real)/$Q$(imaginary) plane at zero gate voltage. (b) Same fit rom (a) plotted for the transmission amplitude. (c) Same fit rom (a) plotted for the microwave phase. }
        \label{fig:circle}
\end{figure*}

\subsection{Supplementary Note 1: \subtxt{L}{ref} design}

We opt for an RF-SQUID architecture to minimize losses to the \gls{cptr}, compared to a DC-SQUID. We exploit the kinetic inductance $\subtxt{L}{kin} \simeq \SI{35}{\nano\henry/\square}$ of the \gls{al} film to build an inductor \subtxt{L}{ref} with dimensions of $\SI{4}{\micro\meter} \times \SI{62}{\micro\meter}$. From this geometry, we estimate $\subtxt{L}{ref} \simeq \SI{0.7}{\nano\henry}$, which corresponds roughly to the inductance of a \gls{jj} with $\subtxt{I}{c} \simeq \SI{480}{\nano\ampere}$---much higher than what was used in the DC-\gls{squid}. The design of \subtxt{L}{ref} accommodates for several competing considerations: we maintain $\subtxt{L}{ref} < \subtxt{L}{r}$ (the inductance of the resonator) so that the impedance of the \gls{squid} remains smaller than the resonator impedance \subtxt{Z}{r}; a thinner arm might interfere with the grounding of the \gls{cpw}; given the thickness of the \gls{al} film, flux focusing will play a major role, meaning the actual \subtxt{L}{ref} may result in a value much higher than designed, yet we must still ensure that $\subtxt{L}{ref} < \subtxt{L}{JJ}(\subtxt{\phi}{ext})$; and finally, based on the circuit in Figure~4(a), we estimated an upper bound for \subtxt{L}{ref} such that for a hypothetical $\subtxt{I}{c,JJ} \simeq \SI{10}{\pico\ampere}$ in the \gls{sc}-\gls{qd}-\gls{sc}, we still achieve a modulation of the resonance frequency \subtxt{f}{r} higher than the \gls{fwhm} for a \gls{cpw} with $\subtxt{Q}{l} \simeq 5 \times 10^{3}$.